\documentclass[epj,a4paper]{svjour}
\usepackage{graphics}
\usepackage{graphicx}
\usepackage{multirow}
\usepackage{comment}
\usepackage{color}
\usepackage{enumitem}
\newcommand{\fig}[1]{fig.~\ref{#1}}
\newcommand{\tab}[1]{table~\ref{#1}}
\newcommand{\Sec}[1]{sec.~\ref{#1}}
\newcommand{\Ref}[1]{ref.~\cite{#1}}
\newcommand{\Refs}[1]{refs.~\cite{#1}}
\newcommand{\mwee}{\,m\,w.\,e. }
\newcommand{\mwe}{\,m\,w.\,e.}
\begin{document}
\title{Cosmic-ray induced background intercomparison with actively shielded HPGe detectors at underground locations}
\author{
T.~Sz\"ucs\inst{1,2,}\thanks{t.szuecs@hzdr.de} \and
D.~Bemmerer\inst{1}\and
T.\,P.~Reinhardt\inst{3}\and
K.~Schmidt\inst{1,3}\and
M.\,P.~Tak\'acs\inst{1,3}\and
A.~Wagner\inst{1}\and
L.~Wagner\inst{1,3}\and
D.~Weinberger\inst{1}\and
K.~Zuber\inst{3}
}
\institute{
Helmholtz-Zentrum Dresden-Rossendorf (HZDR), Dresden, Germany\and
Institute for Nuclear Research (MTA Atomki), Debrecen, Hungary\and
Institute for Nuclear  and Particle Physics, Technische Universit\"at Dresden, Dresden, Germany
 }
\date{Received: date / Revised version: date}
%
\abstract{
The main background above 3\,MeV for in-beam nuclear astrophysics studies with $\gamma$-ray detectors is caused by cosmic-ray induced secondaries. The two commonly used suppression methods, active and passive shielding, against this kind of background were formerly considered only as alternatives in nuclear astrophysics experiments. In this work the study of the effects of active shielding against cosmic-ray induced events at a medium deep location is performed. Background spectra were recorded with two actively shielded HPGe detectors. The experiment was located at 148\,m below the surface of the Earth in the Reiche Zeche mine in Freiberg, Germany. The results are compared to data with the same detectors at the Earth's surface, and at depths of 45\,m and 1400\,m, respectively.
\PACS{
      {26.}{Nuclear astrophysics}\and
      {29.30.Kv}{X- and $\gamma$-ray spectroscopy}\and
      {29.40.Wk}{Solid-state detectors}
     } 
} 
\authorrunning{T. Sz\"ucs \it et al.}
\titlerunning{Background intercomparison with actively shielded HPGe detectors at underground locations}
\maketitle
\section{Introduction}
\label{sec:intro}

In low-energy nuclear astrophysics experiments with high purity germanium (HPGe) detectors, the signal is usually small compared to the laboratory background. Therefore, it is crucial to reduce the background as much as possible. Below 3\,MeV, the $\gamma$ rays from the environmental long lived radionuclides especially from the uranium and thorium decay chains dominate the background, and can be suppressed by a shielding from high density material surrounding the detector. This background region below 3\,MeV is important in activity counting experiments. The background count rate in this energy region depends on the environment, the shielding geometry and the purity of the detection system. The background reduction methods applied there are discussed in details in the literature \cite[and references therein]{Heusser95-ARNPS,Povinec08-APS}. A detailed discussion of this topic is beyond the scope of this paper.

Typically, for in-beam nuclear astrophysics studies $\gamma$ rays with energies higher than 3\,MeV have to be detected. This holds for example for helium burning reactions (e.g. $^{12}$C($\alpha$,$\gamma$)$^{16}$O \cite{Buchmann06-NPA}; $^{16}$O($\alpha$,$\gamma$)$^{20}$Ne \cite{Costantini10-PRC}), for CNO cycle reactions (e.g. $^{14}$N(p,$\gamma$)$^{15}$O \cite{Marta11-PRC}; $^{15}$N(p,$\gamma$)$^{16}$O \cite{Bemmerer09-JPG}), or for reactions important for the production of neutron for the s-process (e.g. $^{18}$O($\alpha$,$\gamma$)$^{22}$Ne \cite{Dababneh03-PRC}). A more precise study of them is needed for nuclear astrophysics purposes \cite{Adelberger11-RMP}.

These and other nuclear astrophysics related studies need background much lower than reachable at the surface \cite{Broggini10-ARNPS}. 
However, these background requirements are not as stringent as those for "silent" experiments (dark matter searches, e.g. \cite{Agostini14-EPJC}). 
Owing to these facts, the LUNA Collaboration operating the world's only underground accelerator for nuclear astrophysics studies was very successful with a background level dominated by the remaining depth-independent neutron flux of about 4\,cm$^{-2}$s$^{-1}$ \cite{Arneodo99-NCSIF}. The scope of this paper is the optimization of the background in $\gamma$-ray detectors for nuclear astrophysics purposes.

In the energy region above 3\,MeV, cosmic-ray induced secondaries (mainly muons) are the dominating background component at the surface of the Earth. As energetic charged particles, the muons are losing their energy by direct ionization depositing up to several hundred MeV in the detector medium.

There are two common methods, active and passive shielding, to reduce the number of recorded background events in the cosmic-ray dominated energy region:

In case of an active shield two detectors are used in anticoincidence. As an anticosmic veto usually a proportional chamber or a scintillator paddle is placed above the experimental setup. From the signal of this detector, an anticoincidence gate is formed and applied to the signal recording chain of the main detector used for the experiment. Since muons are passing through the whole setup causing signals in both detectors, the recorded muon signal can thus be effectively suppressed. Depending on the relative geometry of the veto detector and the investigated energy region, the background count rate with active shielding is a factor of a few tens to a few hundreds lower than without veto.
In other cases, a so-called escape suppression is used. There, an annular veto detector surrounds the detector of interest, usually as closely as possible. In this case, not only the cosmic-ray induced events can be suppressed, but the Compton continuum of the background and/or reaction $\gamma$-rays, too. In this case, the scattered $\gamma$-ray is detected in the veto detector, and a Compton count from the detector of interest can be excluded. This is especially useful when the signal sits on top of the Compton continuum of a higher lying $\gamma$ peak in the spectrum. In our studies HPGe detectors have been used together with surrounding bismuth germanate (BGO) active shielding.

Alternatively, passive shielding may be used against the cosmic rays. With this method the actual number of muons and other cosmic-ray induced particles passing through the detector is reduced, and not only the recorded signal counts. This is the preferable solution, because in this case random coincidences may not decrease the signal. The difficulty is the actual implementation, since cosmic-ray muons have high energy and high penetrability. To passively reduce them to a negligible level for nuclear astrophysics purposes comparable to the ambient depth independent neutron background, requires a shielding equivalent to at least $1500-2000$\,m of water. Practically the overburden of $500-700$\,m deep underground laboratory fulfills this requirement. The exact shielding depth depends on the density of the rock hosting the laboratory. Henceforth in this study the meter water equivalent (m\,w.\,e.) depth unit is used.

The passive shielding method against cosmic-ray induced muons is successfully applied by the LUNA collaboration. The LUNA accelerator is situated in the Gran Sasso National Laboratory (LNGS) in Italy. The laboratory is shielded from cosmic rays by 3800\mwee rock overburden, helping measure cross sections lower than ever reached before \cite{Broggini10-ARNPS,Lemut06-PLB,Anders14-PRL}.

The combination of the two methods is used in rare event search experiments \cite{KlapdorKleingrothaus03-AP,Agostini14-EPJC}, which usually expect their signals in the low energy range.
The effect of the combination of the two methods in the energy range above 3\,MeV  for nuclear astrophysics experiments was only recently investigated experimentally \cite{Szucs10-EPJA,Szucs12-EPJA}. Deep underground, where the muons are negligible compared to other background components, it was found that active shielding causes no additional reduction of the high energy laboratory background \cite{Szucs10-EPJA}. 

At the surface of the Earth, even if active shielding is applied, the remaining background is two orders of magnitude higher than that deep underground \cite{Szucs10-EPJA}. However, shallow underground ($\approx$\,100\mwe), where muons are still the main background source, using an active shield, the background count rate above 3\,MeV was found to become comparable to the one at deep underground ($>$\,1000\mwe) \cite{Szucs12-EPJA}.

The aim of this work is to better understand the effect of the interplay of active shielding and underground settings by extending the available dataset to a medium deep site ($\approx$\,400\mwe) with the same detector used before, and with a second HPGe detector with a thicker active BGO shielding.

\section{Detectors used and sites investigated}
\label{sec:det_and_sites}

\subsection{Review of previous measurements}
\label{sec:prevoius}

The effect of the combination of active and a passive shielding was investigated previously at the Earth's surface at Helmholtz-Zentrum Dresden-Rossendorf (HZDR), at the Felsenkeller laboratory (110\mwe) in Germany \cite{Kohler09-ARI}  and at LNGS (3800\mwe) in Italy. The detailed description of these previous measurements is found in \Refs{Szucs10-EPJA,Szucs12-EPJA}.

\begin{figure*}
\center
  \includegraphics[height=50mm]{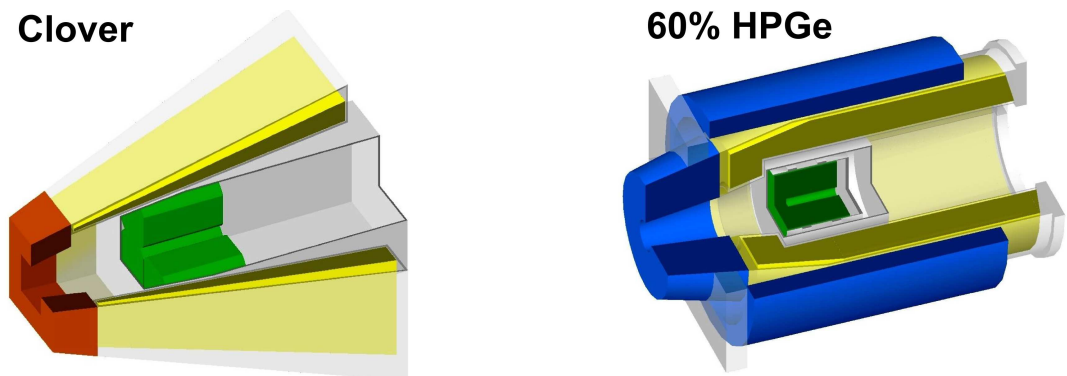}
\caption{Schematic view of the detectors used in this study. The HPGe crystals are green (in case of the Clover the forth crystal is left out for clarity), the BGO crystals are yellow. In case of the Clover the "heavy met" collimator is orange, in case of the 60\%\,HPGe the lead collimator and shield are blue. The end-caps and holding elements are shown in gray.}
\label{fig:det}
\end{figure*}

In these measurement campaigns \cite{Szucs10-EPJA,Szucs12-EPJA} one and the same Clover-type HPGe detector with 122\% relative efficiency, equipped with a BGO veto shielding \cite{Elekes03-NIMA} was used (\fig{fig:det}, left side). The Clover's BGO has a truncated pyramidal shape with 0.3\,cm thickness at the front and 2\,cm thickness at the rear. At the front end of the BGO there is a "heavy met" collimator. "Heavy met" is an alloy of tungsten ($>90$\%), nickel and copper. The collimator has a square shaped 3.5\,cm\,$\times$\,3.5\,cm opening. The same detector was subsequently transported to each site, and laboratory background spectra have been recorded \cite{Szucs10-EPJA,Szucs12-EPJA}. The same detector is again used here.

It was found previously \cite{Szucs10-EPJA,Szucs12-EPJA}, that an additional 5\,cm lead shield surrounding the whole detector has no measurable effect on the high energy background count rate if active shielding is applied. Therefore in the present comparison the previously recorded unshielded spectra are used, and the newly obtained spectra were recorded also without additional lead shield.

\subsection{New measurement in the "Reiche Zeche"	}
\label{sec:new_meas}

The Clover detector has been transported to the Reiche Zeche mine in Freiberg, Germany. The Freiberg Mining Field is an ore deposit of precious and non-ferrous metals in the lower Eastern Ore Mountains in Saxony, Germany. 
The first discovery of silver ore dates from 1168. The first confirmed mining activity at Reiche Zeche dates back to 1384 \cite{Wagenbreth86-book}. The mine is currently used as a teaching, research and visitor mine by TU Bergakademie Freiberg. A possible use as a national underground laboratory has been proposed \cite{Mischo14-Freiberg}.
The present measurement has been done in the so-called Klimakammer 148\,m below the surface. On this level, a former $\gamma$-ray measurement concentrating on the low energy background had been performed in the 1980's \cite{Hebert86-NIMB}.

Beside the Clover, a second HPGe detector with 60\%\, relative efficiency was transported to the same site (hereafter 60\%\,HPGe). This detector is equipped with an annular BGO shield (\fig{fig:det}, right side). The crystals of this BGO have a different shape than the one of the Clover, and are approximately 3\,cm thick, leading to a higher veto efficiency. Around the BGO there is a 2\,cm thick lead shield and at the front a 7\,cm thick lead collimator with a cylindrical opening of 3\,cm diameter to suppress the overall count rate of the veto detector. This is necessary to reduce false veto signals caused by random coincidences.
Subsequently, background spectra have also been recorded with the 60\%\,HPGe at Felsenkeller and at HZDR.

 In table~\ref{tab:sites} the depth of the sites investigated, and detectors used in the comparison are summarized. The histograms have been stored on a daily basis and list mode data have also been recorded to keep track of possible gain changes, which were finally found to be negligible. There was no observable change in the background rate, so in the final analysis the sum of the daily spectra has been used.
 
\begin{table}[t]
\caption{Summary of the sites and detectors used in recent and in previously published studies \cite{Szucs10-EPJA,Szucs12-EPJA}. See \Sec{sec:new_meas} for a detailed description of the highlighted new site and detector.}
\label{tab:sites}
\center
\resizebox{0.99\columnwidth}{!}{
\begin{tabular}{lccc}
\hline\noalign{\smallskip}
\multirow{2}{*}{Site} & Depth & \multicolumn{2}{c}{Recorded energy range} \\
 & [\mwe] & 122\%\,Clover & \textbf{60\%\,HPGe}\\
\noalign{\smallskip}\hline\noalign{\smallskip}
HZDR & 1 & $0.9-74$\,MeV \cite{Szucs12-EPJA} &  $0.3-39$\,MeV\\
Felsenkeller & 110 & $0.3-22$\,MeV \cite{Szucs12-EPJA} & $0.3-39$\,MeV\\
\textbf{Reiche Zeche} & \textbf{400} & $0.5-73$\,MeV & $0.3-41$\,MeV\\
LNGS & 3800 & $0.1-8$\,MeV \cite{Szucs10-EPJA} & no data\\
\noalign{\smallskip}\hline
\end{tabular}
}
\end{table}

\section{Expected effect of the cosmic-ray induced particles on the HPGe detector background based on the literature and on simulations}
\label{sec:disc}

Our study concentrates on the high energy background ($E_\gamma$\,$>$\,3\,MeV), where the natural radioactivity of the surroundings of the setup and of the detector materials plays a negligible role at the surface of the Earth. 
Only this energy region above 3\,MeV is discussed in the following, where  the background is originating either from cosmic-ray induced events or the ambient neutron background.

\subsection{Cosmic-ray induced background above 3\,MeV}
\label{sec:background}

The primary cosmic rays entering the atmosphere of the Earth are light nuclei with very high energy up to $10^{20}$\,eV \cite{Grieder01-book}. In the atmosphere, these particles lose energy via electromagnetic and nuclear processes generating secondary cosmic-ray particles reaching the surface, and penetrating into the Earth's crust. These are mainly muons (hard component), neutrons, protons (nucleonic component), electrons, positrons and gamma rays (soft component).

\begin{figure}[t]
\center
\resizebox{0.99\columnwidth}{!}{
  \includegraphics{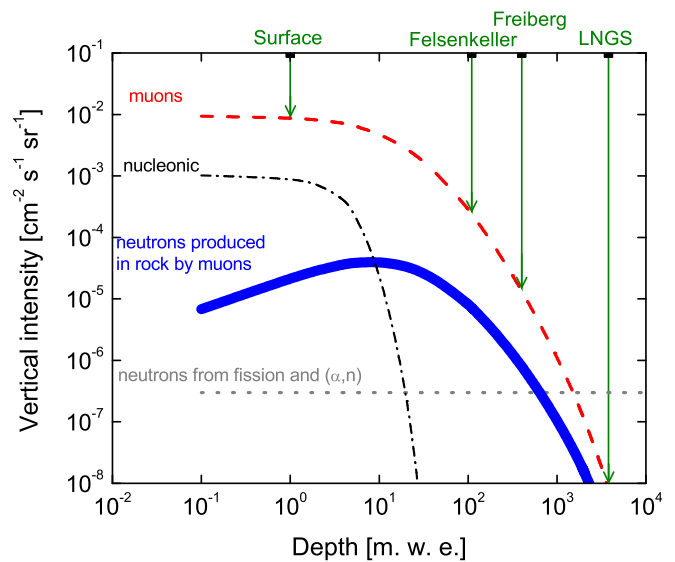}
}
\caption{Qualitative behavior of muon and neutron intensity dependence from the depth based on \cite{Grieder01-book} as described in the text. The typical ($\alpha$,n) neutron flux in Gran Sasso \cite{Arneodo99-NCSIF} is also shown. The depth of the sites investigated in this work are marked by vertical arrows.}
\label{fig:intensity}
\end{figure}

The approximate dependence  of the muon and neutron intensity on the depth is shown in \fig{fig:intensity}, and a qualitative discussion will be given here following \Refs{Grieder01-book,Mei06-PRD}.
\begin{description}[leftmargin=0mm]
\item[\it Red dashed curve:] The depth dependence of the muon intensity, as approximated by eq. (4.35) from \Ref{Grieder01-book}. 
\item[\it Black dot-dashed curve:] The flux of the nucleonic component of the cosmic rays has an attenuation length of about 200\,$\mathrm{g\,cm^{-2}}$ \cite{Heusser95-ARNPS}. The plotted curve is this exponential decrease, converted into vertical intensity for comparison by the Gross transformation (eq. (1.45) in \Ref{Grieder01-book}).
\item[\it Blue solid curve:]  The muon induced neutron production in the rock was calculated as follows using data from \Ref{Mei06-PRD}. The average muon energy is increasing with depth, and the neutron production yield does the same. To take into account this effect, the neutron production yield, shown in table~IV of \Ref{Mei06-PRD} and here in \tab{tab:neutron_prod} in the appendix, was taken as a function of depth, and fitted with a power law (\fig{fig:fit}), similar to the one of fig.~8. of \Ref{Mei06-PRD}. The resulting curve was multiplied by the muon flux, and scaled to match the total produced neutron flux below 1500\mwe, as parametrized by eq.~(13) of \Ref{Mei06-PRD}. After the scaling, the neutron flux curve plotted here and the parametrized curve from \Ref{Mei06-PRD} are within 10\% in the region of $1500-6000$\mwee The resulting curve follows the muon intensity decrease, except at depths smaller than 10\mwee This initial backbending is caused by the smaller amount of material above the given depth, thus smaller amount of production target for muon induced neutrons.
\item[\it Gray dotted curve:]  The typical neutron flux from the environmental radioactivity at Gran Sasso \cite{Arneodo99-NCSIF}. Assuming isotropic total flux, the value was divided by 4$\pi$ to obtain the differential flux as plotted for the other components.
This background is site dependent \cite{Formaggio04-ARNPS}, its flux related to the composition of the rock surrounding the experimental site. However, due to the presence of muon induced neutrons at Felsenkeller and at Reiche Zeche depth, the expected sensitivity of the present setup is not enough to check for a possible site dependence of the ($\alpha$,n) neutron flux at these two sites.
\end{description}

The aim of \fig{fig:intensity} is only to show the general trends of the various components. A precise construction of these curves is beyond the scope of this article. However, the combined uncertainty of the approximations used here is always below a factor of 2.

\subsection{Simulation of the detector response to different cosmic-ray components at the Earth's surface}
\label{sec:simulation}

In order to investigate the detector response to the different components of the cosmic rays, the 60\%\,HPGe and its BGO shield were coded  in a GEANT4 \cite{Agostinelli03-NIMA} simulation (\fig{fig:det}). The cosmic-ray shower library (CRY) \cite{cry} was used as a generator of the comic-ray particles. Six types of secondary are generated with their default momentum, direction and energy distribution (matching the observed quantities). Depending on the type of the secondary, those were classified to be hard component (muons, pions), soft component (electrons, gammas) or nucleonic component (protons, neutrons) prior to the simulation of their interaction with the detection system. The simulation took into account the production of secondary particles (electrons, positrons, neutrons) and their interactions with the detector materials and shielding. With the simulation, also the detector response can be classified into the responses to the different cosmic-ray components, which are attenuated very differently by the rock overburden (\fig{fig:intensity}).
The simulation was only performed for the surface setting. A precise simulation of the underground muon flux and the muon induced neutrons are beyond the scope of this paper and would be computationally very expensive \cite{Kudryavtsev03-NIMA}.

\begin{figure*}[t]
\center
\resizebox{0.99\linewidth}{!}{%
  \includegraphics{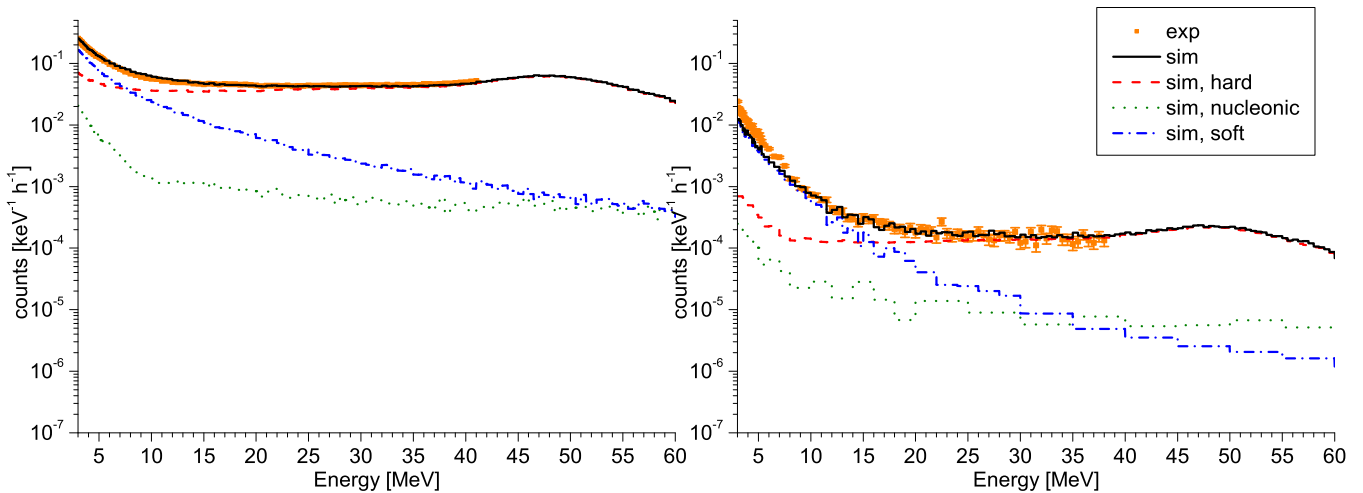}
}
\caption{Measured (orange points with error bars) and simulated (black solid line) laboratory background spectra and its components (hard: red dashed lines, soft: blue dot-dashed lines, nucleonic: green dotted lines) in the 60\%\,HPGe detector. Non suppressed (escape suppressed) spectra are shown in the left (right) panel. The experimental data extended to 41\,MeV (39\,MeV), respectively.}
\label{fig:60HPGe_sim}
\end{figure*}

The simulation (\fig{fig:60HPGe_sim}) reproduces the measured surface spectra within 10\% in the whole energy region under investigation here.
For the vetoed spectra the following conditions have been considered. Every time when an event deposited energy above a given threshold (in our case 75\,keV) in the BGO, the signal of the germanium was vetoed. The threshold was approximated by the experimental constant fraction discriminator (CFD) threshold used in the signal chain of the veto gate forming.
In addition 0.35\% veto loss was introduced to match the experimental spectrum. This amount of lost veto signal can occur in the screening layers between the BGO crystals, or in the light connection between the scintillators and the photo tubes, or in the signal creation of the photo tubes, all of which were neglected in the simulation.

Since the muons are usually not stopped in the detector material, the detector response depends on the detector geometry, mainly on the height of the detectors. The so-called muon peak is formed in the spectrum \cite{Heusser95-ARNPS,Povinec08-APS}. The energy of this broad peak corresponds to the muon energy loss multiplied by the main path length of the muons in the active material. Given the diameter of the 60\%\,HPGe detector (characteristic vertical size), the muon peak should be around 48\,MeV. The simulation reproduces this feature very well.

The simulated spectra divided into their components are shown in \fig{fig:60HPGe_sim}. The effect of the active veto on the various components are discussed in the following.

The active veto is the most efficient against the hard component of the cosmic rays, reducing it by a factor of $200-300$.

Since the 60\%\,HPGe detector has only a 2\,cm thick lead shielding around the BGO, the soft component is also sizable. It even dominates the detector background at lower energies. The energy deposit of the soft component of the cosmic rays drops rapidly with increasing energy. The suppression of the soft component is less efficient than that of the muons, only a factor of $30-50$ (increasing with increasing energy). Therefore, in the escape suppressed spectrum the soft component's relative contribution is even higher. 

The nucleonic component (mainly neutrons) has less than about 5\% of the muon contribution in the non vetoed spectrum and less than about 15\% in the escape suppressed spectrum. Furthermore, it reaches a flat distribution above 10\,MeV in both cases. The active shielding shows rather uniform suppression against this component, about a factor of $50-80$ in the whole energy range.

\subsection{Expected detector response at the underground sites}
\label{sec:signal}

According to \fig{fig:intensity} the main background source at each site, except deep underground, is the muons. The energy deposited by them, and therefore the background spectrum observed by the detectors becomes flat above $10-15$\,MeV up to around 40\,MeV (see simulated spectra in \fig{fig:60HPGe_sim} and experimental in \fig{fig:all}).

By using an active shielding, the direct effect of the muons can be suppressed by a factor of $200-300$ (see \Sec{sec:simulation} above). As shown in \fig{fig:intensity}, already 15\mwee cover removes the nucleonic component of the cosmic rays. More importantly, this depth also removes the soft component of the cosmic rays \cite{Heusser95-ARNPS}.
 
The remaining background at shallow ($\approx$\,100\mwe) and medium ($\approx$\,400\mwe) depth is caused by the muon induced neutrons, in addition to the muons. These neutrons have a hard energy spectrum \cite{Mei06-PRD,Kudryavtsev03-NIMA} up to even the GeV region, just as cosmic-ray induced neutrons \cite{Grieder01-book}. At the surface, the cosmic-ray induced neutron flux is about 20\% of the flux of the muons \cite{Grieder01-book}. According to the simulation (\Sec{sec:simulation}) in the escape suppressed setting, their contribution to the background is less than 15\% of the muon contribution. Conservatively considering a similar efficiency of the veto detector against the muon induced neutrons underground as against the cosmic-ray induced neutrons above ground, and taking into account the flux ratio of the muons and muon induced neutrons, their ratio in the background signal can be estimated. The estimation shows that the escape suppressed shallow depth underground signal still consists mainly of muon signals. The contribution of the muon induced neutrons is less than about 3\% in case of the 60\%\,HPGe. 

However, at medium depth the average muon energy is higher, the number of produced neutrons per muon is higher, the escape suppressed background contribution of those neutrons are less than about 5\% of the muons. 
The main difference between shallow ($\approx$\,100\mwe) and medium deep ($\approx$\,400\mwe) underground is the relative contribution of the ($\alpha$,n) neutrons.  These neutrons can interact with the detector material via elastic and inelastic scattering. However, by these processes they can deposit only their full kinetic energy up to 8\,MeV \cite{Mei06-PRD}. 
In addition, they may cause higher energy signals via radiative neutron capture up to about  10\,MeV. The limit is the highest Q-value of neutron capture in the construction material of the shielding or the detector \cite{Ohsumi02-NIMA}. 
The active shielding has no effect against these signals, therefore no count rate difference is expected with and without the BGO escape suppression.
In the escape suppressed spectra at Reiche Zeche (400\mwe) the ($\alpha$,n) signal rate is expected to become comparable to the muon signal rate, since there are one order of magnitude less muons present than at Felsenkeller (110\mwe) considering similar veto efficiency against the muons at both sites.

Deep underground ($>$\,1000\mwe) the ($\alpha$,n) neutrons have the highest flux among the background components shown in \fig{fig:intensity}. As those can not produce signals higher than 10\,MeV, a drop is expected in the spectrum at this energy.
With the Clover detector the spectra were recorded only up to 8\,MeV, therefore, we can not see this effect in our measurement. However, in \Ref{Bemmerer05-EPJA} a long background recording shows only few counts above 10\,MeV in a very large (57\,kg) BGO detector at LUNA, consequently a practically empty HPGe spectrum is expected above 10\,MeV deep underground (for the purpose of nuclear astrophysics).

Table~\ref{tab:background_sources} summarizes the above discussion, the main background contributors are shown on the different sites in different energy regions. In case of the escape suppressed spectra, the cosmic-ray induced neutrons and muon induced neutrons have a higher contribution to the overall count rate, but are still secondary as discussed in the previous paragraphs. 
As shown previously \cite{Heusser95-ARNPS} the direct cosmic-ray induced neutrons are reduced to negligible level by a few tens of \mwee deep overburden. At a depth of a few hundred \mwee in an escape suppressed settings, the ($\alpha$,n) signal rate is already sizable compared to the muon induced neutron signals in the energy region below 10\,MeV. Deep underground  the signal from ($\alpha$,n) reactions is expected to be the dominant contribution.
\begin{table*}[t]
\caption{Estimated main background sources in different energy regions at different sites. In parentheses the sources of secondary importance are shown. Sources with negligible impact are left out of the table.}
\label{tab:background_sources}
\center
\begin{tabular}{l c |c c |c c c c}																										
\hline\noalign{\smallskip}																									
\multirow{2}{*}{Site}  & Depth &	 \multicolumn{2}{c|}{free running} & \multicolumn{2}{c}{escape suppressed} \\
  & [\mwe] &	$<$\,10\,MeV &  $>$\,10\,MeV & $<$\,10\,MeV &  $>$\,10\,MeV \\
\noalign{\smallskip}\hline\noalign{\smallskip}																						
HZDR 		& 1	& $\mu$; (cosmic n)		& $\mu$; (cosmic n)	& $\mu$; (cosmic n)			& $\mu$; (cosmic n)		\\
Felsenkeller 	& 110	& $\mu$; ($\mu$,n)		& $\mu$; ($\mu$,n)& $\mu$; ($\mu$,n)			& $\mu$; ($\mu$,n)		\\
Reiche Zeche& 400	& $\mu$; ($\mu$,n)		& $\mu$; ($\mu$,n)& $\mu$, $\alpha$,n; ($\mu$,n)	& $\mu$; ($\mu$,n)	\\
LNGS 		& 3800& $\alpha$,n 		& -- -- --		& $\alpha$,n 			& -- -- --\\
\noalign{\smallskip}\hline																									
\end{tabular}																											
\end{table*}

\section{Experimental results and discussion}
\label{sec:results}

\begin{figure*}[t]
\center
\resizebox{0.99\linewidth}{!}{%
  \includegraphics{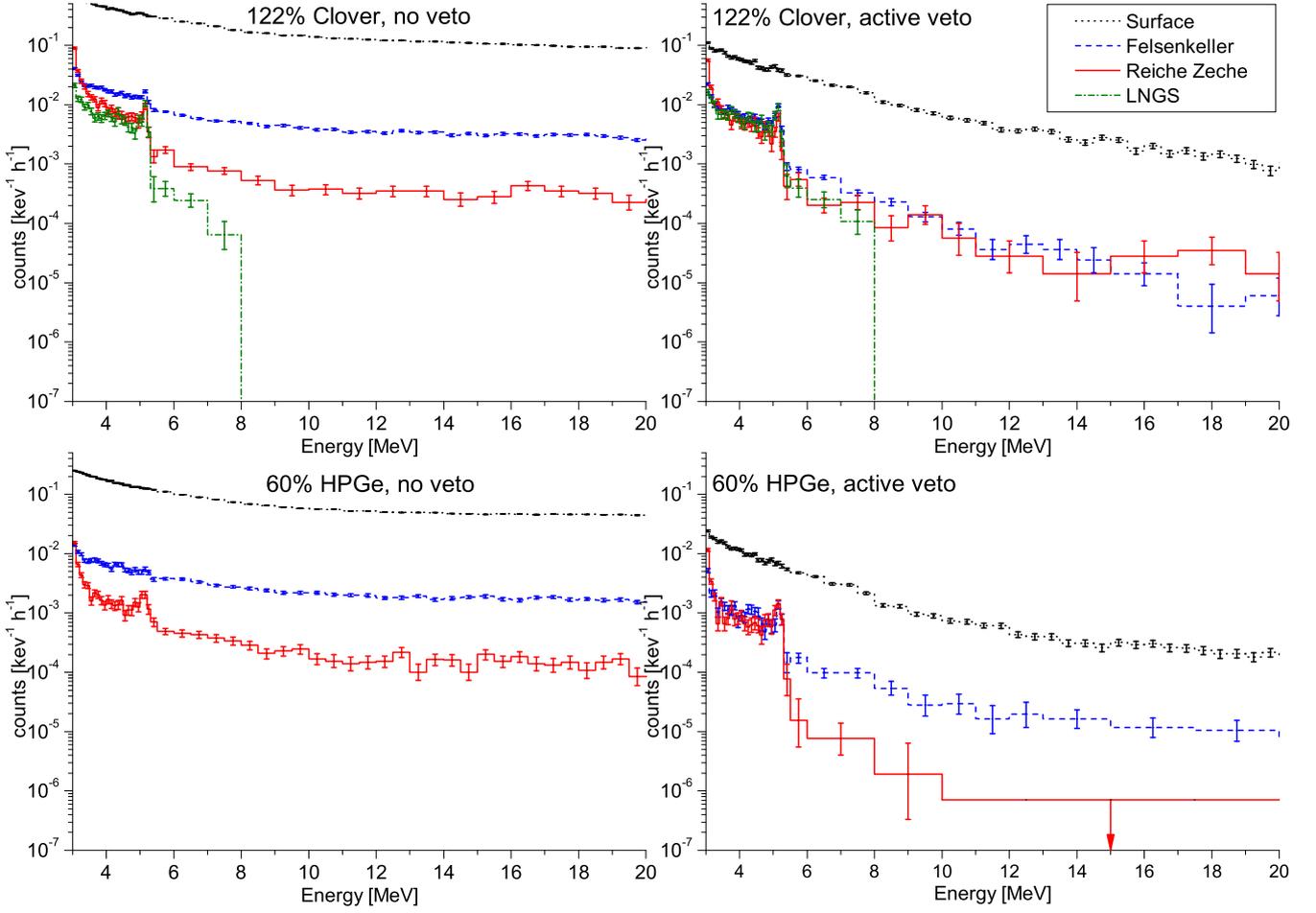}
}
\caption{Comparison of all the spectra recorded. Upper left: Clover, no veto; upper right: Clover, active veto; lower left: 60\%\,HPGe, no veto; lower right: 60\%\,HPGe, active veto. Black dotted lines: surface; blue dashed lines: Felsenkeller (110\mwe); red solid lines: Reiche Zeche (400\mwe); green dash-dotted line: LNGS (3800\mwe) \cite{Szucs10-EPJA}. Numerical values of count rates are presented in \tab{tab}.}
\label{fig:all}
\end{figure*}

Laboratory background spectra were recorded with different dynamical ranges of the preamplifier-amplifier system. A vertical drop in the spectra (\fig{fig:all}) indicates where the recorded spectrum ends, for example at LNGS 8\,MeV is the end of the recorded Clover spectrum \cite{Szucs10-EPJA}.

In the raw histograms, no structure was observed which requires the HPGe resolution. Therefore for a better understanding and clearer view of the results, the spectra have been rebinned.

In the underground spectra a depth independent peak structure is observable around 5.3\,MeV in both HPGe detectors (\fig{fig:all}). It is caused by intrinsic $\alpha$-emitters from a $^{210}$Po contamination of the solder of the crystal. Such a contamination is typical for detectors not specified as ultra low background \cite{Brodzinski87-NIMA}. In order to exclude this effect, the discussion of the data is limited to energies above 5.3\,MeV.
\begin{table*}[!t]
\caption{Comparison of the recorded count rates in different energy regions, in the two detectors, at the investigated sites. The shown numbers are in 10$^{-3}$ counts / (keV hour). When less than 20 counts was observed in an energy bin, Poisson error is quoted and shown instead of the square root of the counts. If no count was observed in a given region "0.000" and also Poisson error of zero count is quoted. The energy regions where no data are available is indicated by "-- -- --".}
\label{tab}
\center
\resizebox{0.99\linewidth}{!}{																																			
\begin{tabular}{@{}l@{\,}l|r@{\,}c@{\,}l r@{\,}c@{\,}l r@{\,}c@{\,}l r@{\,}c@{\,}l | r@{\,}c@{\,}l r@{\,}c@{\,}l r@{\,}c@{\,}l r@{\,}c@{\,}l}																	\multicolumn{2}{c}{} & \multicolumn{12}{c|}{no active shield} & \multicolumn{12}{c}{veto detector active} \\					
\noalign{\smallskip}\hline\noalign{\smallskip}
Detector	&	Site	&	\multicolumn{3}{c}{6--8\,MeV}					&	\multicolumn{3}{c}{8--10\,MeV}					&	\multicolumn{3}{c}{10--15\,MeV}					&	\multicolumn{3}{c|}{15--20\,MeV}					&	\multicolumn{3}{c}{6--8\,MeV}					&	\multicolumn{3}{c}{8--10\,MeV}					&	\multicolumn{3}{c}{10--15\,MeV}					&	\multicolumn{3}{c}{15--20\,MeV}					\\
\noalign{\smallskip}\hline\noalign{\smallskip}
	&	HZDR	&	219	&	$\pm$	&	1	&	154.7	&	$\pm$	&	1.0	&	122.1	&	$\pm$	&	0.5	&	97.8	&	$\pm$	&	0.5	&	20.6	&	$\pm$	&	0.4	&	9.0	&	$\pm$	&	0.2	&	3.89	&	$\pm$	&	0.10	&	1.53	&	$\pm$	&	0.06	\\
122\%	&	Felsenkeller	&	5.74	&	$\pm$	&	0.11	&	4.44	&	$\pm$	&	0.09	&	3.47	&	$\pm$	&	0.05	&	3.01	&	$\pm$	&	0.05	&	0.46	&	$\pm$	&	0.03	&	0.180	&	$\pm$	&	0.019	&	0.044	&	$\pm$	&	0.006	&	0.008	&	$\pm$	&	0.003	\\
Clover	&	Reiche Zeche	&	0.83	&	$\pm$	&	0.08	&	0.45	&	$\pm$	&	0.06	&	0.33	&	$\pm$	&	0.03	&	0.32	&	$\pm$	&	0.03	&	0.21	&	$\pm$	&	0.04	&	0.11	&	$^+_-$	&	$^{0.04}_{0.03}$	&	0.028	&	$^+_-$	&	$^{0.012}_{0.009}$	&	0.028	&	$^+_-$	&	$^{0.012}_{0.009}$	\\
	&	LNGS	&	0.15	&	$\pm$	&	0.03	&	\multicolumn{3}{c}{-- -- --}	&	\multicolumn{3}{c}{-- -- --}	&	\multicolumn{3}{c|}{-- -- --}	&	0.18	&	$^+_-$	&	$^{0.05}_{0.04}$	&	\multicolumn{3}{c}{-- -- --}	&	\multicolumn{3}{c}{-- -- --}	&	\multicolumn{3}{c}{-- -- --}	\\
\noalign{\smallskip}\hline\noalign{\smallskip}
\multirow{3}{*}{\begin{tabular}{@{}l} 60\% \\ HPGe \end{tabular}}
&	HZDR	&	85.8	&	$\pm$	&	0.3	&	62.9	&	$\pm$	&	0.3	&	50.86	&	$\pm$	&	0.17	&	45.79	&	$\pm$	&	0.16	&	3.08	&	$\pm$	&	0.07	&	1.12	&	$\pm$	&	0.04	&	0.479	&	$\pm$	&	0.016	&	0.248	&	$\pm$	&	0.012	\\
	&	Felsenkeller	&	3.19	&	$\pm$	&	0.07	&	2.35	&	$\pm$	&	0.06	&	1.90	&	$\pm$	&	0.04	&	1.72	&	$\pm$	&	0.03	&	0.098	&	$\pm$	&	0.013	&	0.041	&	$\pm$	&	0.008	&	0.020	&	$\pm$	&	0.004	&	0.011	&	$\pm$	&	$0.003$	\\
	&	Reiche Zeche	&	0.40	&	$\pm$	&	0.03	&	0.24	&	$\pm$	&	0.02	&	0.151	&	$\pm$	&	0.011	&	0.146	&	$\pm$	&	0.011	&	0.008	&	$^+_-$	&	$^{0.006}_{0.004}$	&	0.002	&	$^+_-$	&	$^{0.004}_{0.002}$	&	0.000	&	$^+_-$	&	$^{0.001}_{0.000}$	&	0.000	&	$^+_-$	&	$^{0.001}_{0.000}$	\\
\noalign{\smallskip}\hline											
\end{tabular}												
}
\vspace{-2mm}
\end{table*}

\subsection{Background level without active shielding}

The shapes of the experimental spectra without active shielding (\fig{fig:all}, left column) are similar at each measurement site. They all follow the simulated spectrum of the hard component of the cosmic rays (\fig{fig:60HPGe_sim}, red dashed line). The average counting rate decreases with increasing depth. In case of the Clover /60\%\,HPGe/ the count rate is at Felsenkeller (110\mwe) a factor of 38 /27/ less than that at the surface. At Reiche Zeche (400\mwe) there is a reduction of a factor of about 264 /215/. At LNGS (3800\mwe) the counting rate is reduced by a factor of 1460. All reductions have been calculated for the energy range of $6-8$\,MeV, important for the $^{14}$N(p,$\gamma$)$^{15}$O and $^{12}$C($\alpha$,$\gamma$)$^{16}$O reactions.
Similar factors are observable in the other energy ranges above 5.3\,MeV. However, the muon flux reductions at these three sites are about a factor of 50, 600, 5\,$\times$\,$10^6$, respectively. The background reduction is therefore not as strong as the muon flux reduction. 

This noticeable difference clearly shows the increasing relative weight of the depth independent background components at the underground sites to the generally decreasing background counting rate.

\subsection{Background level with active shielding and veto efficiency}
The overall shape of the actively shielded  spectra (\fig{fig:all}, right column) are different from those without veto. There is a steeper decrease in the counting rate with increasing energy.

In case of the Clover, as it was observed previously \cite{Szucs12-EPJA}, already at Felsenkeller (110\mwe) the counting rate is close to that at LNGS (3800\mwe). At Reiche Zeche (400\mwe), the counting rate is consistent with the LNGS level in the range of $6-8$\,MeV.
However, at LNGS no Clover data are available above 8\,MeV. From a LaBr$_3$ spectrum \cite{Szucs12-EPJA} and a spectrum recorded by a BGO detector \cite{Bemmerer05-EPJA} we expect decreasing background up to 10\,MeV, and an almost background free spectrum above this energy (for the purposes of nuclear astrophysics).

At Reiche Zeche the Clover veto seems to be less efficient. This is caused by the depth independent ($\alpha$,n) neutron signals, which are unaffected by the active shielding. Therefore, when the ($\alpha$,n) contribution became sizable the averaged veto efficiency drops, even though the veto efficiency remains the same against the other background components.

As it was previously noted \cite{Szucs10-EPJA}, at LNGS the spectra with and without active shielding are consistent. At deep underground the dominating background component cannot be vetoed.

\subsection{Differences between the two detectors}

In case of the 60\%\,HPGe, a lower background rate is observed with respect to the Clover. In this detector the veto still reduces the background by a similar factor than that at the surface or shallow underground. This suggests, that in this detector the ($\alpha$,n) neutrons produce less signal. There was no single count in the vetoed spectra in Reiche Zeche above 10\,MeV during 11 days of data taking, therefore only an upper limit is quoted  (\fig{fig:all}, lower right).

The different behavior of the two detectors can be attributed to three factors:
\begin{enumerate}
\item difference in the active volume
\item thickness of the BGO shield
\item extra 2\,cm lead shielding in case of the 60\%\,HPGe (see \fig{fig:det}).
\end{enumerate}
\vspace{-1mm}
The non vetoed count rate in the Clover detector is about a factor of two higher, as expected from the larger crystal size.
Beside the higher efficiency, the collimator of the Clover contains tungsten. This material has a much higher radiative neutron capture cross section than lead \cite{Chadwick11-NDS}, enhancing the ($\alpha$,n) signal in the detector.  Below 10\,MeV, the signal from ($\alpha$,n) neutrons is also slightly higher in the Clover compared to the 60\%\,HPGe, because the thinner Clover BGO is less efficient as a passive neutron shield, and the BGO of the 60\%\,HPGe is surrounded by additional 2\,cm of lead.

Also the Clover has lower veto efficiency against the muon induced neutrons, due to its thinner BGO. The number of veto signals created by these neutrons in the BGO scales with the active volume of the veto detector.

\subsection{Consequences for low-energy nuclear astrophysics experiments}

In a previous work based on the observed counting rates with the Clover detector at Felsenkeller (110\mwe), detailed counting rate estimates were shown for number of astrophysically relevant reactions with a single HPGe detector setup \cite{Szucs12-EPJA}.

The present data confirm the previous finding  \cite{Szucs12-EPJA} that for number of astrophysical scenarios, already shallow-underground accelerators will enable significant progress. However, it is still true that the background counting rate keeps decreasing with depth \cite{Laubenstein04-ARI}, so that deep-underground sites are needed to push the counting rate limit even lower. In the previous example of the $^{12}$C($\alpha$,$\gamma$)$^{16}$O case \cite{Szucs12-EPJA}, in one representative setup a lower center-of-mass energy limit of 835\,keV was reached shallow underground, as opposed to 720\,keV deep underground.

Some recommendations can be given for shallow-un\-der\-ground HPGe detectors for accelerator-based experiments:
\begin{enumerate}
\item Use a large solid angle active veto detector.
\item Use thick, high density material for the veto (e.g. 3\,cm of BGO).
\item Surround the active shield by an additional lead shield.
\item Avoid materials with high radiative neutron capture cross section like tungsten.
\end{enumerate}

\section{Summary and outlook}
\label{sec:outlook}
 
The laboratory background has been measured at a depth of 148\,m by one and the same detector, with which spectra were recorded and published previously. Data exists from the Earth's surface, from a shallow underground laboratory (45\,m) \cite{Szucs12-EPJA} and from deep underground (1400\,m) \cite{Szucs10-EPJA}. These measurements, except the deep underground one, have also been repeated by another escape suppressed detector.

It is found that by using an active shielding already at 400\mwee the depth independent neutron background is the dominant background component if an active shielding is applied. Therefore, an equivalent background to deep underground is reached for $E_\gamma$\,$>$\,3\,MeV.

In nuclear astrophysics motivated studies already a background level found here is sufficient. In this kind of experiments the signal count rate is so low that even without background the lowest energy points take a few month of running time to reach sufficient precision.

Since these direct reaction parameter studies of astrophysical interest require long measuring times, there is a need which was also expressed in the present NuPECC Long Range Plan 2010 \cite{NuPECC} for nuclear physics in Europe, that the installation of one or more higher-energy underground accelerators is strongly recommended.
The findings of this paper opens the way for medium and shallow underground laboratories to also host particle-accelerators for such studies, speeding up progress in the field.

As a result of these findings in the Felsenkeller tunnel system a 5\,MV pelletron accelerator is currently under installation \cite{Szucs12-JPCS,Bemmerer15-PoS}.

\section*{Acknowledgments}
The authors thank Detlev Degering (VKTA) for his help during the data acquisition in the Fel\-sen\-kel\-ler. The support by the staff of the Reiche Zeche mine, TU Bergakademie Freiberg, is gratefully acknowledged.
The project was supported by the Helmholtz Association through the Nuclear Astrophysics Virtual Institute (NAVI, HGF VH-VI-417), and by OTKA (K101328, K108459).

\begin{appendix}
\section*{Appendix}
\begin{table}[!!!h]
\caption{Muon induced neutron production rates.}
\label{tab:neutron_prod}
\center
\resizebox{0.99\columnwidth}{!}{
\begin{tabular}{l c c r@{\,$\times$\,}l}
\hline\noalign{\smallskip}
\multirow{2}{*}{Reference} & Depth & $\left<E_\mu\right>$ & \multicolumn{2}{c}{$\left<n\right>$} \\
 & [\mwee]  & [GeV] & \multicolumn{2}{c}{[$n/($$\mu$ $g\,cm^{-2})]$} \\
\noalign{\smallskip}\hline\noalign{\smallskip}
Hertenberger \cite{Hertenberger95-PRC}	& 20		& 13	&  $(2.0\,\pm\,0.7)$ 		& 10$^{-5}$ \\
Bezrukov \cite{Bezrukov73-SJNP}		& 25 		& 14.7&  $(4.7\,\pm\,0.5)$	& 10$^{-5}$ \\
Boehm  \cite{Boehm00-PRD}			& 32		& 16.5&  $(3.6\,\pm\,0.3)$	& 10$^{-5}$ \\
Bezrukov \cite{Bezrukov73-SJNP}	 	& 316 		& 55	&  $(1.21\,\pm\,0.12)$	& 10$^{-4}$ \\
Enikeev \cite{Enikeev87-SJNP} 			& 750 		& 120	&  $(2.15\,\pm\,0.15)$	& 10$^{-4}$ \\
Corrected LVD data \cite{Aglietta99-axiv,Mei06-PRD} 	& 3100	& 270	&  $(4.5\,\pm\,1.2)$ 	& 10$^{-4}$ \\
Aglietta (LSD) \cite{Aglietta89-NCSIF}				& 5000	& 346	&  $(5.3\,\pm\,1.1)$ 	& 10$^{-4}$ \\
\noalign{\smallskip}\hline
\end{tabular}
}\vspace{-5mm}
\end{table}
\begin{figure}[!!h]
\center
\resizebox{0.99\linewidth}{!}{%
  \includegraphics{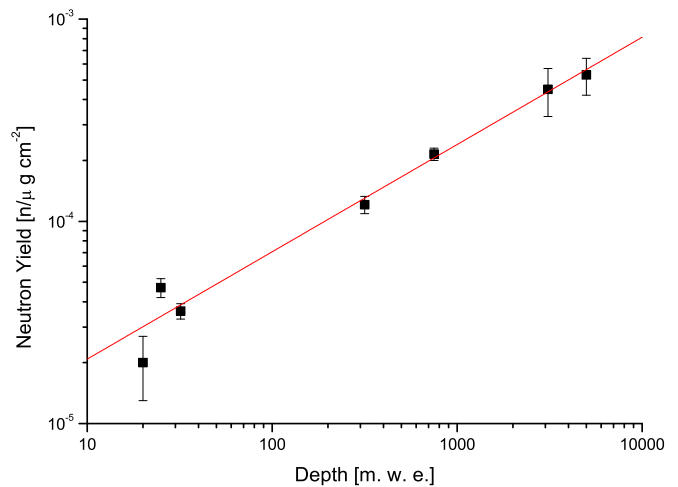}
}
\caption{Neutron production rate as a function of depth fitted with a power law.}
\label{fig:fit}
\end{figure}

\end{appendix}

\end{document}